\documentclass[11pt,a4paper]{article}

\usepackage{amsfonts}
\usepackage{amsmath}
\usepackage{amssymb}
\usepackage{amsthm}
\usepackage{fullpage}
\usepackage{mleftright}
\usepackage{enumitem}
\usepackage{color}
\usepackage[normalem]{ulem}
\usepackage{cancel}
\usepackage[f]{esvect}
\usepackage{hyperref}

\DeclareMathOperator*{\E}{\mathbb{E}}

\DeclareMathOperator\supp{supp}

\newcommand*{\heavy}{\mathrm{Heavy}}
\newcommand*{\subheavy}{\mathrm{Heavy}}
\newcommand*{\norm}[1]{\left\|#1\right\|}
\newcommand*{\dist}{\mbox{dist}}
\newcommand{\R}{\mathbb{R}}

\newcommand{\F}{\mathbb{F}}
\newcommand*\cupdot{\mathbin{\mathaccent\cdot\cup}}
\newcommand{\res}[2]{\mathrm{res}_{#2\leftarrow#1}}

\numberwithin{equation}{section}
\newtheorem{theorem}{Theorem}[section]
\newtheorem*{theorem*}{Theorem}
\newtheorem{lemma}[theorem]{Lemma}

\newtheorem{proposition}[theorem]{Proposition}

\theoremstyle{definition}
\newtheorem{definition}[theorem]{Definition}
\newtheorem*{remark*}{Remark}

\mleftright
\begin{document}

\title{Improved Small Set Expansion in High Dimensional Expanders}
\author{
	Tali Kaufman\thanks{Bar-Ilan University, Israel. Email: \texttt{kaufmant@mit.edu}.} \and
	David Mass\thanks{The Academic College of Tel Aviv-Yaffo, Israel. Email: \texttt{dudimass@gmail.com}.}}
\maketitle

\begin{abstract}
Small set high dimensional expansion over general sheaves has recently became central to several major breakthroughs. These include the construction of classical locally testable and quantum LDPC codes with constant rate and linear distance, $\Omega(n)$-levels lower bounds for the Sum-of-Squares hierarchy, and the proof of the NLTS conjecture.

In this work, we improve upon the state-of-the-art results for small set expansion in simplicial complexes, for both constant and general sheaves. Compared to one line of work (e.g., [KM22, DD24]), we improve upon their result by obtaining \emph{strong expansion} for small sets. Compared to another line of work (e.g., [KKL14, EK16, KM21, FK22]), we get an \emph{exponential improvement} in the size of sets for which expansion is guaranteed.

Our result is based on a novel technique that bridges these two lines of work via a ``local vs. global'' case analysis: for ``local'' sets, which are concentrated within small links, we adapt ideas from the ``fat machinery'' of [KKL14, EK16]; for ``global'' sets, which are spread across the complex, we exploit global averaging operators in the spirit of [KM22, DD24].
\end{abstract}
\newpage

\section{Introduction}
\subsection{Background}
\paragraph{High dimensional expansion.}
High dimensional expansion is a generalization of expansion in graphs to high dimensional objects. In this work we focus on \emph{simplicial complexes}, which can be thought of as hypergraphs that are downwards closed. Specifically, A $d$-dimensional simplicial complex $X$ is a $(d+1)$-hypergraph with the property that if $\sigma \in X$ and $\tau \subset \sigma$ then also $\tau \in X$. A hyperedge of size $k+1$ is called a \emph{$k$-face} of the complex.

We briefly describe the two main notions of high dimensional expansion, which generalize graph expansion to higher dimensions. The first, \emph{local spectral expansion}, is concerned with the spectral expansion of the underlying graph of each local piece of the complex. These local pieces are called \emph{links}, where the link of a face $\sigma \in X$ is the subcomplex 
obtained by taking all faces $\tau \supseteq \sigma$ and removing $\sigma$ from them.

The other notion, generalizing edge expansion in graphs, is called \emph{coboundary} or \emph{cosystolic expansion}. For the sake of introduction, we describe this notion only over $\F_2$.\footnote{See \S\ref{sec:coboundary-expansion} for the general definition.} Consider a \emph{$k$-cochain} $f : X(k) \to \F_2$, which is a $\{0,1\}$-assignment to the $k$-faces of the complex. The \emph{coboundary} of $f$ is a $\{0,1\}$-assignment to the $(k+1)$-faces defined by $\delta f(\sigma) = \sum_{v \in \sigma}f(\sigma \setminus \{v\})$ for every $\sigma \in X(k+1)$. The coboundary of $f$ can be thought of as a set of unsatisfied constraints, where each $(k+1)$-face corresponds to a parity check of its $k$-subfaces. A $k$-cochain that satisfies all the constraints is called a \emph{$k$-cocycle}, and a $k$-cochain that is obtained as the coboundary of a $(k-1)$-cochain is called a \emph{$k$-coboundary}.

Coboundary and cosystolic expansion\footnote{The difference between coboundary and cosystolic expansion is whether there exist cocycles that are not coboundaries, i.e., satisfying assignments which are not obtained by the coboundary of an assignment of one dimension lower. For the sake of introduction, we can think of them as similar notions.} are concerned with the relationship between the fraction of unsatisfied constraints and the distance of the cochain from a satisfying assignment. More formally, a complex is said to be a $\beta$-cosystolic expander in dimension $k$ if for every $k$-cochain $f$,
\begin{equation}\label{intro-cosys-def}
    \norm{\delta f} \ge \beta \min_{g\,:\,\delta(g)=\mathbf{0}}\norm{f-g},  
\end{equation}
where $\norm{\cdot}$ is the normalized hamming weight.

For example, when $X = (X(0),X(1))$ is a (connected) graph, this notion coincides exactly with edge expansion. It can be seen by mapping a set of vertices $S \subseteq X(0)$ to its indicator function $\mathbf{1}_S$. Then $\delta(\mathbf{1}_S) = \mathbf{1}_{E(S,\overline{S})}$ is exactly the indicator function of the set of edges leaving $S$. In this case, the only functions $g$ that satisfy $\delta(g) = \mathbf{0}$ are the constant functions. Hence, $\beta$-cosystolic expansion in dimension $0$ can be equivalently written as
$$\forall S \subset X(0), S \ne \emptyset: \quad
\frac{\|E(S,\overline{S})\|}{\min\{\norm{S}, \|\overline{S}\|\}} =
\frac{\|\delta(\mathbf{1}_S)\|}{\min\{\norm{\mathbf{1}_S},\|\mathbf{1}_{\overline{S}}\|\}} \ge \beta.$$

\paragraph{Small set expansion.}
With the previous example in mind, small set expansion in simplicial complexes can be viewed as a generalization of graph small set expansion to higher dimensions. A naive approach, similar to graphs, would require that every cochain with small weight has a proportionally large coboundary. However, this approach fails for $k$-cochains with $k > 0$. For instance, fix an arbitrary vertex $v \in X(0)$ and take the set of edges containing this vertex, i.e., define the $1$-cochain $f : X(1) \to \F_2$ by $f(e) = 1$ for every $e \ni v$, and $f(e) = 0$ for all other edges. $f$ is completely supported on a single vertex, so $\norm{f}$ is small. However, $\norm{\delta f} = 0$, since every triangle contains either zero or two edges that contain $v$. This phenomenon is actually a general property of chain complexes, known as $\delta\circ\delta = 0$. Note that in this example, $f = \delta(\mathbf{1}_{\{v\}})$.

Roughly speaking, this should be the only obstacle to expanding small cochains. To be more precise, a $k$-cochain is said to be \emph{minimal} if its weight cannot be reduced by adding a $k$-coboundary to it. It is said to be \emph{locally minimal} if it is minimal at every proper link, i.e., its weight cannot be reduced by adding a $k$-coboundary that is supported on a lower dimensional face (hence the term \emph{locally} minimal). We say that a complex is a \emph{$(c_1,c_2)$-small set expander}\footnote{In some other works, this property is explicitly termed ``small set locally minimal expansion''. For brevity, we call it here just small set expansion.} in dimension $k$, if for every locally minimal $k$-cochain $f$,
\begin{equation*}
\norm{f} \le c_1 \quad\Rightarrow\quad \norm{\delta f} \ge c_2\norm{f}.
\end{equation*}

\paragraph{On the importance of small set expansion.}
In~\cite{KKL14, EK16}, the authors introduced small set expansion in higher dimensions as a tool for establishing cosystolic expansion.  \cite{KKL14, EK16} have shown that expansion of small locally minimal cochains implies cosystolic expansion in one dimension lower. Interestingly, even \emph{weak expansion} is sufficient for this purpose, i.e., having $\norm{\delta f} > 0$ for every (non-zero) small locally minimal $k$-cochain $f$ implies cosystolic expansion in dimension $k-1$.

A sequence of works followed~\cite{KKL14, EK16} and proved small set expansion for broader sets of coefficients (e.g.,~\cite{KM21, KM22, DD24} for groups, and~\cite{FK22} for sheaves, discussed below) as a method to deduce cosystolic expansion in the same way.

Recently, remarkable breakthrough works constructed locally testable codes~\cite{DEL+22, PK22, FK24} and quantum LDPC codes~\cite{PK22, LZ22, DHLV23} with constant rate and linear distance. Their constructions are based on certain expanding \emph{square complexes} over \emph{sheaves} (discussed in the next paragraph). Following their works, small set expansion has found wider applications in other areas. For instance, \cite{HL22} showed that the construction of~\cite{LZ22} admits small set expansion, which in turn yields hard instances for $\Omega(n)$ levels of the Sum-of-Squares hierarchy, and~\cite{ABN23} proved the NLTS conjecture, again by small set expansion of similar square complexes.

\paragraph{Sheaves.}
Over the past few years, sheaf theory on high dimensional expanders has emerged, providing an essential algebraic framework for these major breakthroughs. Broadly speaking, sheaves generalize the notion of a constant coefficient group (such as $\F_2$ that we discussed above) by assigning a local vector space to each face of the complex. Again, for simplicity, we focus on the case where these vector spaces are over $\F_2$. Formally, a sheaf $\mathcal{F}$ on a simplicial complex $X$ is defined by the following two ingredients:
\begin{enumerate}[label=(\arabic*)]
    \item An $\F_2$-vector space $\mathcal{F}(\sigma)$ for every $\sigma \in X$.
    \item An $\F_2$-linear map $\res{\sigma}{\tau} : \mathcal{F}(\sigma) \to \mathcal{F}(\tau)$ for every $\sigma \subset \tau \in X$, subject to the requirement that $\res{\sigma}{\tau}\circ\res{\rho}{\sigma} = \res{\rho}{\tau}$ for every $\rho \subset \sigma \subset \tau \in X$.
\end{enumerate}

Now, a $k$-cochain $f$ assigns an element of $\mathcal{F}(\sigma)$ to every $\sigma \in X(k)$, i.e., $f \in \prod_{\sigma \in X(k)}\mathcal{F}(\sigma)$. The coboundary operator is defined as before, but it has to take into account the linear maps as well. Formally, for any $\tau \in X(k+1)$,
\begin{equation*}
    \delta f(\tau) = \sum_{\substack{\sigma \in X(k) \\ \sigma \subset \tau}}\res{\sigma}{\tau}f(\sigma).
\end{equation*}

With the coboundary operator in hand, the rest of the definitions (e.g., cosystolic, coboundary, and small set expansion) remain the same. In this language, defining $\mathcal{F}(\sigma) = \F_2$ for every $\sigma \in X$, and $\res{\sigma}{\tau} = \mathrm{id}$ for every $\sigma \subset \tau \in X$, recovers the definitions we discussed at the beginning of the introduction.

\subsection{Our contribution}
In this work, we improve upon the state-of-the-art results for small set expansion in simplicial complexes, for \emph{both constant and general sheaves}.

\begin{theorem}[Main theorem, informal]\label{intro-main-thm}
    Let $X$ be a $d$-dimensional simplicial complex and $\mathcal{F}$ a sheaf on $X$. If $X$ is a local spectral expander and its links are $\beta$-coboundary expanders, then for any $0 < k < d$, $X$ is a $(c_1,c_2)$-small set expander in dimension $k$,
    where\footnote{The $\simeq$ notation hides factors that depend on the local spectral expansion of the complex, which is assumed to be arbitrarily small. See Theorem~\ref{thm-small-set-exp} for the exact constants.}
    $$c_1 \simeq \frac{\beta^k}{(k+2)!} \quad\mbox{and}\quad c_2 = \frac{\beta^k}{k!(k+1)^4}.$$
\end{theorem}

At a high level, previous works can be divided into two main paradigms. One line of work~\cite{KKL14, EK16, KM21, FK22} uses what has been termed ``fat machinery'' to establish small set expansion. However, this framework is significantly affected by the weight of the cochain, consequently obtaining expansion only for cochains of \emph{very small} weight. Technically, they establish $(c_1,c_2)$-small set expansion in dimension $k$, where $c_1 \lesssim (\beta^k/k!)^{2^k}$.

The other line of work~\cite{KM22, DD24} uses local-to-global averaging operators, which improve upon the weight of cochains that expand but at the cost of the expansion quality. More precisely, they show that every locally minimal \emph{$k$-cocycle} $f$ satisfies $\norm{f} \gtrsim \beta^k/(k+1)!$, which in turn implies that any (non-zero) locally minimal cochain of weight $\norm{f} \lesssim \beta^k/(k+1)!$ is not a cocycle, i.e., it satisfies $\norm{\delta f} > 0$. In our notation, they improve upon the constant $c_1$, but provide no guarantee for $c_2$.

In this work, we obtain ``the best of both worlds'' and show strong expansion for locally minimal cochains of weight $\lesssim \beta^k/(k+2)!$. We achieve an exponential improvement in the weight of expanding cochains over the first line of work, and improve upon the expansion quality of the second line of work.


\paragraph{Our technique.}
In order to describe our technique, let us briefly discuss the strategies of previous works. The first line of work~\cite{KKL14, EK16, KM21, FK22} can be thought of as deducing the expansion of small locally minimal cochains from their local expansion in the links. Their analysis divides faces of smaller dimensions into ``fat faces'' and ``non-fat faces''. Their main argument (with a few variations between the works) builds upon the observation that whenever a cochain contains many fat faces, their local coboundary as seen in the links is actually a global coboundary. The main caveat of this framework is that to benefit from these fat faces, the cochain must be \emph{very} small.

The second line of work~\cite{KM22, DD24} can be thought of as deducing expansion from the global structure of the cochain. They utilize (with a few variations between the works) certain local-to-global averaging operators, which are applicable only to locally minimal \emph{cocycles}. In this way, they obtain a larger lower bound on the weight of a locally minimal cocycle, which in turn implies only $\norm{\delta f} > 0$ for any small locally minimal $f$, i.e., with no guarantee on the expansion quality. The main caveat of this approach is the dependency on the global structure of the cochain being a cocycle.

Our result is based on a novel technique that bridges these two frameworks via a ``local vs. global'' case analysis.

\textbf{Key idea:} We argue that any cochain either has a ``local'' structure in the sense that it is concentrated within links of some smaller dimension, or it has a ``global'' structure in the sense that it looks much like a cocycle. In the local case, we adapt ideas from the ``fat machinery'' approach of the first link of work, and in the global case, we exploit the global averaging operators in the spirit of the second line of work.

\paragraph{Decoupling spectral and coboundary expansion.}
Interestingly, our technique decouples the spectral expansion from the coboundary expansion in the links. In the local case, we deduce expansion solely from coboundary expansion of the links, and in the global case, we deduce expansion relying only on spectral arguments. This stands in contrast to all previous works, which simultaneously utilized both spectral and coboundary expansion of the links throughout their proofs.

\subsection{Discussion}
\paragraph{Applying our technique to general posets.}
We improve upon the state-of-the-art results for small set expansion in simplicial complexes, for both constant and general sheaves. However, as mentioned above, recent breakthroughs rely not only on expansion over sheaves, but also on complexes which are not simplicial, e.g., square complexes. A natural direction is to investigate whether our ``local vs. global'' technique introduced in this work can be applied to other types of complexes, potentially improving upon the aforementioned results.

\paragraph{More applications for small set expansion in simplicial complexes.}
As far as we are aware, for most applications of small set expansion in simplicial complexes (e.g., as a tool for proving cosystolic expansion), having $\norm{\delta f} > 0$ for every (non-zero) small locally minimal cochain is sufficient. However, the expansion quality plays a crucial role in recent applications that are derived from square complexes. For instance, it affects the number of levels that cannot be refuted by the Sum-of-Squares hierarchy~\cite{HL22}, and it plays a central role in the proof of the NLTS conjecture~\cite{ABN23}. It would be interesting to explore whether this type of strong expansion in simplicial complexes yields further applications.

Specifically, in a recent work,~\cite{DLV24} constructs ``almost good'' quantum LTCs, using expanding cubical complexes over sheaves. They suggest that a possible direction toward obtaining ``good'' quantum LTCs is to venture away from cubical complexes and use sheaves on simplicial complexes instead (see the discussion in~\cite{DLV24}). It would be interesting to see whether our improved small set expansion over sheaves can be useful toward this goal.

\subsection{Organization}
In Section~\ref{sec:preliminaries} we provide required preliminaries. In Section~\ref{sec:small-set-expansion} we prove our strong small set expansion theorem. Propositions~\ref{prop-exp-from-heavy-faces} and~\ref{prop-exp-from-non-heavy-faces} demonstrate our case analysis of general cochains: If the cochain is ``local'', i.e., it is concentrated within links of some smaller dimension, we get expansion from Proposition~\ref{prop-exp-from-heavy-faces}, and if the cochain is ``global'', i.e., it looks like a cocycle, we get expansion from Proposition~\ref{prop-exp-from-non-heavy-faces}. We prove our main theorem assuming these two propositions and then we turn to prove each proposition.

As this work bridges between two lines of work, some lemmas with slight variations appear in previous works and some others appear with different notations. We comment through the work regarding the similarities and differences with previous works.

\section{Preliminaries}\label{sec:preliminaries}
\subsection{Simplicial complexes}\label{sec:simplicial-complexes}
A pure $d$-dimensional simplicial complex $X$ is a downward closed $(d+1)$-hypergraph, namely, $X$ is a collection of sets (hyperedges) of size $d+1$ together with all of their subsets. A hyperedge of size $k+1$ is called a \emph{$k$-face}, and the set of all $k$-faces of $X$ is denoted by $X(k)$. For instance, $X(0)$, $X(1)$, and $X(2)$ correspond to the sets of vertices, edges, and triangles, respectively. We also assume that every complex contains a unique $(-1)$-face which corresponds to the empty set, i.e., $X(-1) = \{\emptyset\}$. All complexes in this work are assumed to be pure and simplicial.

For any $k$-face $\sigma \in X(k)$, the \emph{link} of $\sigma$ is a $(d-k-1)$-dimensional complex defined by $X_\sigma = \{\tau \setminus \sigma \mid \sigma \subseteq \tau \in X\}$, i.e., the subcomplex obtained by taking all faces $\tau \supseteq \sigma$ and removing $\sigma$ from them.

We also consider \emph{ordered} faces of $X$, where an ordered $k$-face is a $(k+1)$-tuple $(v_0,\dots, v_k)$ such that $\{v_0,\dots, v_k\} \in X(k)$. The set of all ordered $k$-faces in $X$ is denoted by $\vv{X}(k)$. We assume that the complex comes with an arbitrary ordering on its vertices, so there is a one-to-one correspondence between the unordered $k$-faces and the ordered $k$-faces of the complex. For convenience, when it is clear from the context, we might refer to $\sigma \in X$ as both an unordered face and its unique corresponding ordered face. When we wish to emphasize that a face is ordered, we denote it by $\vec{\sigma}$.

Let $\vec{\sigma} = (v_0, \dots, v_\ell)$ and $\vec{\tau} = (v_{\ell+1}, \dots, v_k)$ be two disjoint ordered faces. We denote by $\vv{\sigma\tau} = (v_0, \dots, v_k)$ the ordered $k$-face that is obtained by the concatenation of $\vec{\sigma}$ and $\vec{\tau}$. For any $0 \le i \le \ell$, we denote by $\vec{\sigma_i} = (v_0, \dots, v_{i-1}, v_{i+1}, \dots, v_\ell)$ the ordered $(\ell-1)$-face that is obtained by removing $v_i$ from $\vec{\sigma}$.

\paragraph{Weights and probabilities.}
Let $X$ be a pure $d$-dimensional simplicial complex. We assume that $X$ comes with a probability distribution over its $d$-faces $P_d : X(d) \to [0,1]$. Consider the following random process: choose a random $d$-face $\sigma^{(d)}$ according to $P_d$, and then, iteratively for $k = d-1, \dots, -1$, remove a uniformly random vertex from $\sigma^{(k+1)}$ to obtain $\sigma^{(k)}$. This random process naturally induces a probability distribution $P_k : X(k) \to [0,1]$ over the $k$-faces for every $k < d$, where for any $\sigma \in X(k)$,
\[
P_k(\sigma) = \Pr_{\sigma^{(d)},\dots,\sigma^{(-1)}}[\sigma^{(k)} = \sigma].
\]

For any set of $k$-faces $S \subseteq X(k)$, we define its weight by
\[
\norm{S} = \Pr_{\sigma^{(d)},\dots,\sigma^{(-1)}}[\sigma^{(k)} \in S]
= \Pr_{\sigma \in X(k)}[\sigma \in S],
\]
where the probability distribution at the right-hand side (and throughout this work) is taken according to $P_k$.

For any $\ell$-face $\sigma \in X(\ell)$, the link $X_\sigma$ inherits the probability distribution from $X$ by conditioning on $\sigma$, i.e., for any $\ell \le k \le d$ and any $\tau \in X_\sigma(k-\ell-1)$,
\[
P^{X_\sigma}_{k-\ell-1}(\tau) = \Pr_{\sigma^{(d)},\dots,\sigma^{(-1)}}[\sigma^{(k)} = \sigma \cup \tau \mid \sigma^{(\ell)} = \sigma].
\]

For any set of $k$-faces $S \subseteq X(k)$ and any $\ell < k$, the following decomposition of $S$ into links of $\ell$-faces follows immediately from the law of total probability:
\[
\norm{S} = \sum_{\sigma \in X(\ell)}\Pr_{\sigma^{(d)},\dots,\sigma^{(-1)}}[\sigma^{(k)} \in S \mid \sigma^{(\ell)} = \sigma] \cdot P_\ell(\sigma) =
\E_{\sigma \in X(\ell)}\left[\Pr_{\tau \in X_\sigma(k-\ell-1)}[\sigma \cup \tau \in S]\right].
\]

\subsection{Spectral expansion}
A weighted graph is called a $\lambda$-one-sided spectral expander if the second largest eigenvalue of its normalized adjacency matrix is upper bounded by $\lambda$.

A $d$-dimensional simplicial complex $X$ is said to be a $\lambda$-one-sided local spectral expander if for every $k \le d-2$ and every $\sigma \in X(k)$, the underlying graph of $X_\sigma$\footnote{The weighted graph whose vertices are $X_\sigma(0)$ and edges are $X_\sigma(1)$.} is a $\lambda$-one-sided spectral expander.

We will use the following standard expander mixing lemma, which first appeared in~\cite{AC88}:
\begin{lemma}[Expander mixing lemma]\label{many-outgoing-edges}
    Let $G=(V,E)$ be a $\lambda$-one-sided spectral expander. For any set of vertices $S \subseteq V$,
    $$\norm{E(S)} \le \norm{S}^2 + \lambda\norm{S},$$
    where $E(S)$ is the set of edges with both endpoints in $S$.
\end{lemma}

\subsection{Sheaves}
Let $X$ be a $d$-dimensional simplicial complex and $\F$ a field.\footnote{One can similarly define sheaves over abelian groups or over $R$-modules for a ring $R$. For simplicity, we consider only sheaves over fields, which are called $\F$-sheaves; see~\cite{FK24} for general definitions and examples.} A sheaf $\mathcal{F}$ on $X$ consists of:
\begin{enumerate}[label=(\arabic*)]
    \item An $\F$-vector space $\mathcal{F}(\sigma)$ for every $\sigma \in X$.
    \item An $\F$-linear map $\res{\sigma}{\tau} : \mathcal{F}(\sigma) \to \mathcal{F}(\tau)$ for every $\sigma \subset \tau \in X$, subject to the requirement that $\res{\sigma}{\tau} \circ \res{\rho}{\sigma} = \res{\rho}{\tau}$ for every $\rho \subset \sigma \subset \tau \in X$.
\end{enumerate}
The maps $\res{\sigma}{\tau}$ are called the \emph{restriction maps} of $\mathcal{F}$.

\paragraph{Sheaf cohomology.}
Let $X$ be a $d$-dimensional simplicial complex and $\mathcal{F}$ a sheaf on $X$. For any $-1 \le k \le d$, a \emph{$k$-cochain} with coefficients in $\mathcal{F}$ is a function $f$ that for every $\vec{\sigma} \in \vv{X}(k)$ assigns an element of $\mathcal{F}(\sigma)$, subject to the requirement that for every $\vec{\sigma} = (v_0,v_1,\dots,v_k) \in \vv{X}(k)$ and every permutation $\pi \in \mbox{Sym}(k+1)$, $$f(\pi\vec{\sigma}) := f((v_{\pi(0)},v_{\pi(1)},\dots,v_{\pi(k)})) = \mbox{sgn}(\pi)f(\vec{\sigma}).$$

The space of all $k$-cochains with coefficients in $\mathcal{F}$ is denoted by $C^k(X,\mathcal{F})$.

For any $-1 \le k \le d-1$, any $k$-cochain $f \in C^k(X,\mathcal{F})$ induces a $(k+1)$-cochain by the coboundary operator $\delta^k : C^k(X,\mathcal{F}) \to C^{k+1}(X,\mathcal{F})$, defined for any $\vec{\sigma} \in \vv{X}(k+1)$ by
\[\delta^k f(\vec{\sigma}) = \sum_{i=0}^{k+1}(-1)^i\res{\sigma_i}{\sigma}f(\vec{\sigma_i}).\]

A direct calculation shows that for any $f \in C^k(X,\mathcal{F})$, indeed $\delta^k f \in C^{k+1}(X,\mathcal{F})$, and also that $\delta^{k+1}\delta^k f = \mathbf{0}$. We usually omit the superscript $k$ from $\delta^k$ and denote the coboundary operator simply by $\delta$, where $k$ is clear from context.

The kernel of $\delta^k$ is called the space of {\em $k$-cocycles} and denoted by
$$Z^k(X,\mathcal{F}) = \mbox{ker}\,\delta^k = \{f \in C^k(X,\mathcal{F}) \mid \delta^k f = \mathbf{0} \}.$$
The image of $\delta^{k-1}$ is called the space of {\em $k$-coboundaries} and denoted by
$$B^k(X,\mathcal{F}) = \mbox{im}\,\delta^{k-1} = \{\delta^{k-1} f \mid f \in C^{k-1}(X,\mathcal{F}) \},$$
with the convention that $\delta^{-2} = 0$.
Since $\delta^{k} \circ \delta^{k-1} = 0$ for every $-1 \le k \le d-1$, it follows that $$B^k(X,\mathcal{F}) \subseteq Z^k(X,\mathcal{F}) \subseteq C^k(X,\mathcal{F}).$$

\subsection{Coboundary and cosystolic expansion}\label{sec:coboundary-expansion}
Let $X$ be a $d$-dimensional simplicial complex and $\mathcal{F}$ a sheaf on $X$.
For any $f \in C^k(X,\mathcal{F})$, the normalized hamming weight of $f$ is defined by
$$\norm{f} = \Pr_{\sigma \in X(k)}[f(\sigma) \ne 0],$$
where the probability is taken according to the distribution as described in \S\ref{sec:simplicial-complexes}.

Since the weight of a cochain is dependent only on its non-zero elements, it is often convenient to consider the set $\supp(f)$, i.e., the set of non-zero elements in $f$, and define equivalently
$$\norm{f} = \norm{\supp(f)} = \Pr_{\sigma \in X(k)}[\sigma \in \supp(f)].$$

The distance between two $k$-cochains $f,g \in C^k(X,\mathcal{F})$ is defined by $\dist(f,g) = \norm{f - g}$.

\paragraph{Cosystolic expansion.}
For any $-1 \le k \le d-1$, $X$ is called a {\em $(\beta,\varepsilon)$-cosystolic expander} in dimension $k$, if the following two conditions hold:
\begin{enumerate}[label=(\arabic*)]
    \item For every $f \in C^k(X,\mathcal{F})\setminus Z^k(X,\mathcal{F})$,
    $$\frac{\norm{\delta(f)}}{\dist(f,Z^k(X,\mathcal{F}))} \ge \beta,$$
    where $\dist(f,Z^k(X,\mathcal{F})) = \min\{\dist(f,g) \;|\; g \in  Z^k(X,\mathcal{F}) \}$.
    \item For every $f \in Z^k(X,\mathcal{F})\setminus B^k(X,\mathcal{F})$, $$\norm{f} \ge \varepsilon.$$
\end{enumerate}

The second condition ensures that there are no small cocycles which are not coboundaries.

\paragraph{Coboundary expansion.}
Coboundary expansion is similar to cosystolic expansion with the extra condition that $Z^k(X,\mathcal{F}) = B^k(X,\mathcal{F})$. Formally,
for any $-1 \le k \le d-1$, $X$ is called a {\em $\beta$-coboundary expander} in dimension $k$, if for every $f \in C^k(X,\mathcal{F})\setminus B^k(X,\mathcal{F})$,
$$\frac{\norm{\delta(f)}}{\dist(f,B^k(X,\mathcal{F}))} \ge \beta,$$
where $\dist(f,B^k(X,\mathcal{F})) = \min\{\dist(f,g) \;|\; g \in  B^k(X,\mathcal{F}) \}$.

\subsection{Locally minimal cochains}
\paragraph{Restricting sheaves to links.}
Let $X$ be a $d$-dimensional simplicial complex and $\mathcal{F}$ a sheaf on $X$. For any $\sigma \in X$, the sheaf $\mathcal{F}$ induces a sheaf $\mathcal{F}_\sigma$ restricted to the link $X_\sigma$, defined by:
\begin{enumerate}[label=(\arabic*)]
    \item $\mathcal{F}_\sigma(\tau) = \mathcal{F}(\sigma \cup \tau)$ for every $\tau \in X_\sigma$, and
    \item $\res{\tau}{\rho}^{\mathcal{F}_\sigma} = \res{\sigma \cup \tau}{\sigma \cup \rho}^{\mathcal{F}}$ for every $\tau \subset \rho \in X_\sigma$.
\end{enumerate}

\paragraph{Localization.}
Let $f \in C^{k}(X,\mathcal{F})$ be a $k$-cochain. For any $0 \le \ell \le k$ and $\sigma \in X(\ell)$, the \emph{localization} of $f$ to the link $X_\sigma$, denoted by $f_\sigma \in C^{k-\ell-1}(X_\sigma,\mathcal{F}_\sigma)$, is a $(k-\ell-1)$-cochain in $X_\sigma$, defined by $f_{\sigma}(\vec{\tau}) = f(\vv{\sigma\tau})$ for every ordered $(k-\ell-1)$-face $\vec{\tau} \in \vv{X_\sigma}(k-\ell-1)$ (where $\vec{\sigma}$ is the unique ordered face corresponding to $\sigma$).

\paragraph{Minimal and locally minimal cochains.}
A $k$-cochain $f \in C^k(X,\mathcal{F})$ is said to be {\em minimal} if its weight cannot be reduced by adding a coboundary to it. Namely, for every coboundary $g \in B^k(X,\mathcal{F})$ it holds that $\norm{f} \le \norm{f-g}$. Recall that the distance of $f$ from the coboundaries is defined by
$\dist(f, B^k(X,\mathcal{F})) = \min \{\norm{f-g} \;|\; g \in B^k(X,\mathcal{F}) \}$. Since $\mathbf{0} \in B^k(X,\mathcal{F})$, it follows that $\norm{f} \ge \dist(f,B^k(X,\mathcal{F}))$ for every $f \in C^k(X,\mathcal{F})$. Hence, $f$ is said to be minimal if and only if $\norm{f} = \dist(f, B^k(X,\mathcal{F}))$.

A $k$-cochain $f \in C^k(X,\mathcal{F})$ is said to be \emph{locally minimal} if the localization of $f$ to the link of $\sigma$ is minimal for every non-empty $\sigma \in X$, i.e., $f_\sigma$ is minimal in $X_\sigma$ for every $\sigma \in X\setminus \{\emptyset\}$.

\section{Small set expansion}\label{sec:small-set-expansion}
We recall the definition of small set expansion from the introduction.
\begin{definition}[Small set expansion]
    Let $X$ be a $d$-dimensional simplicial complex and $\mathcal{F}$ a sheaf on $X$. Let $k \in \{0,\dots,d-1\}$, and $c_1,c_2 \in [0,1]$. $X$ is said to be a $(c_1,c_2)$-small set expander in dimension $k$, if for every locally minimal $k$-cochain $f \in C^k(X, \mathcal{F})$,
    $$\norm{f} \le c_1 \quad\Rightarrow\quad \norm{\delta f} \ge c_2\norm{f}.$$
\end{definition}

Our main theorem is a local criterion for small set expansion.

\begin{theorem}[Main theorem]\label{thm-small-set-exp}
Let $X$ be a $d$-dimensional simplicial complex and $\mathcal{F}$ a sheaf on $X$. Let $k \in \{1,\dots,d-1\}$, $\lambda \in [0,1)$ and $\beta_0,\dots,\beta_k \in (0,1]$. If $X$ is a $\lambda$-one-sided local spectral expander and for every $\sigma \in X(\ell)$, $0 \le \ell \le k$, $X_\sigma$ is a $\beta_\ell$-coboundary expander in dimension $k-\ell-1$, then $X$ is a $(c_1,c_2)$-small set expander in dimension $k$, where
$$c_1 = \left(\frac{\beta_k}{k+2}-\lambda\right)\frac{\beta_0\dotsb\beta_{k-1}}{(k+1)!} - 2\lambda \quad\mbox{and}\quad
c_2 = \frac{\beta_0\dotsb\beta_k}{k!(k+1)^4}\norm{f}.$$
\end{theorem}

\begin{remark*}
    For the sake of comparison with previous works, when $\mathcal{F}$ is a constant sheaf (e.g.,~\cite{KM21, KM22, DD24}), then $\beta_k = 1$. In the case of a general sheaf (e.g.,~\cite{FK22}), $\beta_k$ could be smaller than $1$.
\end{remark*}

The main idea is to deal separately with two types of small cochains. If the cochain has a global structure in the sense that it looks like a cocycle, we would gain expansion by spectral arguments. Otherwise, if the cochain has a local structure in the sense that it is concentrated within links of some smaller dimension, we would gain expansion by the coboundary expansion of these links.

In order to distinguish between these two types of cochains, we provide another way of looking at neighborhoods of faces, in addition to the links.

Let $X$ be a $d$-dimensional simplicial complex, $\mathcal{F}$ a sheaf on $X$, $f \in C^k(X, \mathcal{F})$, and $\sigma \in X(\ell)$, $\ell \le k < d$.
Recall that the link of $\sigma$ is a $(d-\ell-1)$-dimensional simplicial complex defined by $X_\sigma = \{\tau \setminus \sigma \;|\; \sigma \subseteq \tau \in X\}$, and the localization of $f$ to $X_\sigma$ is defined by $f_\sigma(\vec{\tau}) = f(\vv{\sigma\tau})$, for every $\vec{\tau} \in \vv{X_\sigma}(k-\ell-1)$.

We define $Y_\sigma^k$ to be the set of $k$-faces that together with $\sigma$ form a $(k+1)$-face. Formally, \[Y_\sigma^k  := \{\tau \setminus \{v\} \mid v \in \sigma \subset \tau \in X(k+1)\}.\]
We denote by $f^\sigma = f|_{Y_\sigma^k}$ the restriction of $f$ to the $k$-faces in $Y_\sigma^k$.
We also define a probability distribution on $Y_\sigma^k$, $P^{Y_\sigma^k} : Y_\sigma^k \to [0,1]$, induced by the probability distribution of $X$: for every $k$-face $\tau \setminus \{v\} \in Y_\sigma^k$, its probability corresponds to choosing the $(k-\ell)$-face $\tau\setminus\sigma$ in the link $X_\sigma$, and then choosing $v \in \sigma$ uniformly at random. Formally,
\[P^{Y_\sigma^k}(\tau \setminus \{v\}) = \frac{P^{X_\sigma}_{k-\ell}(\tau \setminus \sigma)}{\ell+1},\]
where we recall that $P^{X_\sigma}_{k-\ell}$ is the probability distribution over $(k-\ell)$-faces in the link $X_\sigma$.

The definitions of $Y_\sigma^k$ and $f^\sigma$ are motivated by the following structural observation. Consider a $(k+1)$-face $\tau \supset \sigma$. Note that any $k$-face contained in $\tau$ either contains $\sigma$ or does not contain $\sigma$. Consequently, the $k$-faces contained in $\tau$ can be decomposed into $(k-\ell-1)$-faces in the link $X_\sigma$ and $k$-faces in $Y_\sigma^k$. We note that this observation and the definition of $f^\sigma$ appeared in~\cite{KM22}, and later also in~\cite{DD24} with a different notation.

Informally speaking, we expect that for a cochain $f$ that looks like a cocycle, i.e., a ``global'' cochain, most faces $\sigma$ will satisfy $\norm{f_\sigma} \lesssim \norm{f^\sigma}$. The following definition allows us to distinguish between cochains that are ``global'' and cochains that are concentrated within links of lower dimension, i.e., ``local'' cochains.

\begin{definition}[Heavy faces\footnote{We use the term ``heavy faces'' in order to distinguish it from the notion of ``fat faces'' of previous works. We say that a face $\sigma$ is ``heavy'' if $\norm{f_\sigma}$ is large \emph{relative} to $\norm{f^\sigma}$, whereas a face $\sigma$ is said to be ``fat'' according to previous works, if $\norm{f_\sigma}$ is larger than some \emph{absolute constant}.}]
Let $X$ be a $d$-dimensional simplicial complex, $\mathcal{F}$ a sheaf on $X$, $f \in C^k(X, \mathcal{F})$, and $\sigma \in X(\ell)$, for $0 \le \ell < k \le d-1$. Assume that $X_\sigma$ is a $\beta$-coboundary expander in dimension $k-\ell-1$. We say that $f$ is \emph{heavy} on $\sigma$ if
\[\norm{f_\sigma} > (\ell+2)\beta^{-1}\norm{f^\sigma}.\]
We denote by $\mbox{Heavy}_\ell(f) = \{\sigma \in X(\ell) \mid f \text{ is heavy on } \sigma\}$ the subset of $\ell$-faces that are heavy.
\end{definition}

The two aforementioned cases are dealt in the following two propositions.

\begin{proposition}[Expansion in the local case]\label{prop-exp-from-heavy-faces}
Let $X$ be a $d$-dimensional simplicial complex and $\mathcal{F}$ a sheaf on $X$. Let $0 \le \ell < k \le d-1$, and assume that for every $\sigma \in X(\ell)$, $X_\sigma$ is a $\beta_\ell$-coboundary expander in dimension $k-\ell-1$. Then for any locally minimal $f \in C^k(X,\mathcal{F})$,
\[
\norm{\delta f} \ge \frac{\beta_\ell}{\ell+2}\E_{\sigma \in \heavy_\ell(f)}[\norm{f_\sigma}]\!\cdot\!\norm{\heavy_\ell(f)}.
\]
\end{proposition}

\begin{proposition}[Expansion in the global case]\label{prop-exp-from-non-heavy-faces}
Let $X$ be a $d$-dimensional $\lambda$-one-sided local spectral expander and $\mathcal{F}$ a sheaf on $X$. Let $k \in \{1,\dots, d-1\}$ and assume that for every $\sigma \in X(\ell)$, $0 \le \ell \le k$, $X_\sigma$ is a $\beta_\ell$-coboundary expander in dimension $k-\ell-1$. For any $f \in C^k(X,\mathcal{F})$, if
\[
\norm{f} \le \big(\frac{\beta_k}{k+2}-\lambda\big)\frac{\beta_0\dotsb\beta_{k-1}}{(k+1)!} - 2\lambda,
\]
then
\[\norm{\delta f} \ge \beta_k\norm{f} - \sum_{\ell=0}^{k-1}\frac{(k+1)(k+2)!}{\beta_{\ell+1}\dotsb\beta_{k-1}(\ell+2)!}\E_{\sigma \in \subheavy_\ell(f)}[\norm{f_\sigma}]\!\cdot\!\norm{\heavy_\ell(f)}.
\]
\end{proposition}

We note that the assumption of $\beta_\ell$-coboundary expansion in the links of $\ell$-faces, $0 \le \ell < k$, is not required for the proof of Proposition~\ref{prop-exp-from-non-heavy-faces}, but only for heavy faces to be defined. In other words, one could choose any arbitrary constants $\beta_\ell$ for $0 \le \ell < k$, and Proposition~\ref{prop-exp-from-non-heavy-faces} would still hold.

We first prove Theorem~\ref{thm-small-set-exp} assuming Propositions~\ref{prop-exp-from-heavy-faces} and~\ref{prop-exp-from-non-heavy-faces}, and then prove the propositions.

\begin{proof}[Proof of Theorem \ref{thm-small-set-exp}]
If there exists $0 \le \ell \le k-1$ such that
$$\E_{\sigma \in \subheavy_\ell(f)}[\norm{f_\sigma}]\!\cdot\!\norm{\heavy_\ell(f)} \ge \frac{\beta_{\ell+1}\dotsb\beta_k(\ell+2)!}{k(k+1)(k+2)! + \beta_0\dotsb\beta_{k-1}}\norm{f},$$
then by Proposition~\ref{prop-exp-from-heavy-faces}
$$\norm{\delta f} \ge \frac{\beta_\ell}{\ell+2} \cdot \frac{\beta_{\ell+1}\dotsb\beta_k(\ell+2)!}{k(k+1)(k+2)! + \beta_0\dotsb\beta_{k-1}}\norm{f} \ge \frac{\beta_0\dotsb\beta_k}{k!(k+1)^4}\norm{f},$$
where the second inequality holds since $\beta_\ell \le 1$ for every $\ell \ge 0$.

Otherwise, it holds that
$$\sum_{\ell=0}^{k-1}\frac{(k+1)(k+2)!}{\beta_{\ell+1}\dotsb\beta_{k-1}(\ell+2)!}\E_{\sigma \in \subheavy_\ell(f)}[\norm{f_\sigma}]\!\cdot\!\norm{\heavy_\ell(f)} \le \frac{\beta_k k(k+1)(k+2)!}{k(k+1)(k+2)! + \beta_0\dotsb\beta_{k-1}}\norm{f}.$$
Then by Proposition~\ref{prop-exp-from-non-heavy-faces}
$$
\begin{aligned}
\norm{\delta f} &\ge \beta_k\norm{f} - \sum_{\ell=0}^{k-1}\frac{(k+1)(k+2)!}{\beta_{\ell+1}\dotsb\beta_{k-1}(\ell+2)!}\E_{\sigma \in \subheavy_\ell(f)}[\norm{f_\sigma}]\!\cdot\!\norm{\heavy_\ell(f)} \\[8pt]&\ge
\beta_k\norm{f} - \frac{\beta_k k(k+1)(k+2)!}{k(k+1)(k+2)! + \beta_0\dotsb\beta_{k-1}}\norm{f} \\[8pt]&\ge \frac{\beta_0\dotsb\beta_k}{k!(k+1)^4}\norm{f}.
\end{aligned}
$$
\end{proof}

\subsection*{Proof of Proposition \ref{prop-exp-from-heavy-faces}}
    In this case, when the cochain is concentrated within links of lower dimension $\ell < k$, we show that the local coboundary expansion inside the links of heavy $\ell$-faces implies global expansion.

    Let $f$ be a locally minimal $k$-cochain.
    Consider a heavy $\ell$-face $\sigma \in \mbox{Heavy}_\ell(f)$ and a $(k-\ell)$-face $\tau \in X_\sigma(k-\ell)$ such that $\delta(f_\sigma)(\tau) \ne 0$. Denote the ordered faces corresponding to $\sigma$ and $\tau$ by $\vec{\sigma} = (v_0, v_1, \dots, v_\ell)$ and $\vec{\tau} = (v_{\ell+1}, v_{\ell+2}, \dots, v_{k+1})$. By definition of the local coboundary,
    \[
    0 \ne \delta(f_\sigma)(\vec{\tau}) = \sum_{i=\ell+1}^{k+1}(-1)^{i-\ell-1}\res{\tau_i}{\tau}^{\mathcal{F}_\sigma}f_\sigma(\vec{\tau_i}) = \sum_{i=\ell+1}^{k+1}(-1)^{i-\ell-1}\res{\sigma\cup\tau_i}{\sigma\cup\tau}^{\mathcal{F}}f(\vv{\sigma\tau_i}).
    \]
    Now, if $f((\vv{\sigma\tau})_i) = 0$ for every $0 \le i \le \ell$, then
    \[
    \delta f(\vv{\sigma\tau}) = \sum_{i=0}^{k+1}(-1)^{i}\res{(\sigma\cup\tau)_i}{\sigma\cup\tau}f((\vv{\sigma\tau})_i) =
    (-1)^{\ell+1}\sum_{i=\ell+1}^{k+1}(-1)^{i-\ell-1}\res{\sigma\cup\tau_i}{\sigma\cup\tau}f(\vv{\sigma\tau_i}) \ne 0.
    \]
    Therefore, in this case, a local non-zero coboundary implies a global non-zero coboundary, i.e.,
    \begin{equation}\label{exp-from-heavy-faces-eq-1}
    \delta(f_\sigma)(\vec{\tau}) \ne 0 \;\wedge\; \forall_{0 \le i \le \ell} \;f((\vv{\sigma\tau})_i) = 0 \quad\Rightarrow\quad \delta f(\vv{\sigma\tau}) \ne 0.
    \end{equation}
    
    Let $\sigma \in \heavy_\ell(f)$ be a heavy $\ell$-face. We will show that many $(k-\ell)$-faces in $X_\sigma$ satisfy the hypothesis of \eqref{exp-from-heavy-faces-eq-1}. Recall the probability distribution over $Y_\sigma^k$: 
    the probability of a $k$-face $\rho \in Y_\sigma^k$ is the probability of obtaining $\rho = (\sigma \cup \tau) \setminus \{v\}$ by choosing a $(k-\ell)$-face $\tau \in X_\sigma(k-\ell)$ according to the probability distribution of $X_\sigma(k-\ell)$ and a vertex $v \in \sigma$ uniformly at random. It holds that
    \begin{equation}\label{exp-from-heavy-faces-eq-2}
    \begin{aligned}
    \norm{f^\sigma} &= \Pr_{\substack{\tau \in X_\sigma(k-\ell)\\ i \in \{0,\dots, \ell\}}}[f((\vv{\sigma\tau})_i) \ne 0] \\&=
    \frac{1}{\ell+1}\sum_{i = 0}^\ell\Pr_{\tau \in X_\sigma(k-\ell)}[f((\vv{\sigma\tau})_i) \ne 0] \\[5pt]&\ge
    \frac{1}{\ell+1}\Pr_{\tau \in X_\sigma(k-\ell)}[\exists_{0 \le i \le \ell} \mbox{ s.t. } f((\vv{\sigma\tau})_i) \ne 0].
    \end{aligned}
    \end{equation}
    
    Combining \eqref{exp-from-heavy-faces-eq-1} and \eqref{exp-from-heavy-faces-eq-2} yields
    \begin{equation*}
    \begin{aligned}
    \Pr_{\tau \in X_\sigma(k-\ell)}[\delta f(\vv{\sigma\tau}) \ne 0] &\ge
    \Pr_{\tau \in X_\sigma(k-\ell)}[\delta(f_\sigma)(\vec{\tau}) \ne 0 \;\wedge\; \forall_{0 \le i \le \ell} \;f((\vv{\sigma\tau})_i) = 0] \\[5pt]&\ge
    \Pr_{\tau \in X_\sigma(k-\ell)}[\delta(f_\sigma)(\vec{\tau}) \ne 0] - \Pr_{\tau \in X_\sigma(k-\ell)}[\exists_{0 \le i \le \ell} \mbox{ s.t. } f((\vv{\sigma\tau})_i) \ne 0] \\[5pt]&\ge
    \norm{\delta(f_\sigma)} - (\ell+1)\norm{f^\sigma} \\[5pt]&\ge \beta_\ell\norm{f_\sigma} - \frac{(\ell+1)\beta_\ell}{\ell+2}\norm{f_\sigma} \\&= \frac{\beta_\ell}{\ell+2}\norm{f_\sigma},
    \end{aligned}
    \end{equation*}
    where the first inequality follows by~\eqref{exp-from-heavy-faces-eq-1}, the second inequality follows by the law of total probability, the third inequality follows by~\eqref{exp-from-heavy-faces-eq-2}, and the last inequality follows since $X_\sigma$ is a $\beta_\ell$-coboundary expander in dimension $k-\ell-1$, $f$ is locally minimal, and $\sigma \in \heavy_\ell(f)$.

    It follows that
    \begin{align*}
    \norm{\delta f} &= \E_{\sigma \in X(\ell)}\left[\Pr_{\tau \in X_\sigma(k-\ell)}[\delta f(\vv{\sigma\tau}) \ne 0]\right] \\[5pt]&\ge
    \E_{\sigma \in \subheavy_\ell(f)}\left[\Pr_{\tau \in X_\sigma(k-\ell)}[\delta f(\vv{\sigma\tau}) \ne 0]\right]\!\cdot\!\norm{\heavy_\ell(f)} \\[5pt]&\ge 
    \frac{\beta_\ell}{\ell+2}\E_{\sigma \in \subheavy_\ell(f)}[\norm{f_\sigma}]\!\cdot\!\norm{\heavy_\ell(f)},
    \end{align*}
    which completes the proof.

\subsection*{Proof of Proposition \ref{prop-exp-from-non-heavy-faces}}
In this case, when there are only a few heavy links in every dimension, we can show by spectral arguments that there are many $(k+1)$-faces that contain exactly one $k$-face $\sigma$ for which $f(\sigma) \ne 0$.

The proof of Proposition~\ref{prop-exp-from-non-heavy-faces} follows from the following two lemmas.

\begin{lemma}\label{lem-exp-from-non-heavy-faces}
    Let $X$ be a $d$-dimensional $\lambda$-one-sided local spectral expander and $\mathcal{F}$ a sheaf on $X$. Let $k \in \{1, \dots, d-1\}$ and assume that for every $\sigma \in X(k)$, $X_\sigma$ is a $\beta_k$-coboundary expander in dimension $-1$. Then for any $f \in C^k(X, \mathcal{F})$,
    $$\norm{\delta f} \ge (k+2)\left(\beta_k\norm{f} - (k+1)\E_{\sigma \in X(k)}[\norm{f_\sigma}\norm{f^\sigma}]\right).$$
\end{lemma}

\begin{lemma}\label{lem-bound-near-times-infront}
    Let $X$ be a $d$-dimensional $\lambda$-one-sided local spectral expander and $\mathcal{F}$ a sheaf on $X$. For any $f \in C^k(X, \mathcal{F})$, $0 \le k < d$ and $0 \le \ell \le k$,
    $$\E_{\sigma \in X(\ell)}[\norm{f_\sigma}\norm{f^\sigma}] \le
    \E_{\sigma \in X(\ell-1)}[\norm{f_\sigma}^2] + \lambda\norm{f}.$$
\end{lemma}

We note that both lemmas are inspired from previous works. A slight variation of Lemma~\ref{lem-exp-from-non-heavy-faces} (with regard to constant sheaves) has appeared in~\cite{KM22}, and Lemma~\ref{lem-bound-near-times-infront}, with a different notation, has appeared in~\cite{DD24}. For completeness, we provide proofs for both lemmas after we show that Proposition~\ref{prop-exp-from-non-heavy-faces} follows from them.

\begin{proof}[Proof of Proposition \ref{prop-exp-from-non-heavy-faces}]
    First, by Lemma~\ref{lem-bound-near-times-infront} applied to $\ell=k$,
    \begin{equation}\label{prop-exp-from-non-heavy-faces-eq0}
    \E_{\sigma \in X(k)}[\norm{f_\sigma}\norm{f^\sigma}] \le \E_{\sigma \in X(k-1)}[\norm{f_\sigma}^2] + \lambda\norm{f}.
    \end{equation}
    Then, for any $0 \le \ell < k$,
    \begin{equation}\label{prop-exp-from-non-heavy-faces-eq1}
    \begin{aligned}
    \hspace{-10pt}
    \E_{\sigma \in X(\ell)}[{\norm{f_\sigma}^2}] &=
    \E_{\sigma \in \overline{\heavy_\ell(f)}}[{\norm{f_\sigma}^2}]\!\cdot\!\big\|\overline{\heavy_\ell(f)}\big\| + \E_{\sigma \in \subheavy_\ell(f)}[{\norm{f_\sigma}^2}]\!\cdot\!\norm{\heavy_\ell(f)} \\[5pt]&\le
    \frac{\ell+2}{\beta_\ell}\E_{\sigma \in \overline{\subheavy_\ell(f)}}[{\norm{f_\sigma}}\norm{f^\sigma}]\!\cdot\!\big\|\overline{\heavy_\ell(f)}\big\| + \E_{\sigma \in \subheavy_\ell(f)}[{\norm{f_\sigma}}]\!\cdot\!\norm{\heavy_\ell(f)} \\[5pt]&\le
    \frac{\ell+2}{\beta_\ell}\E_{\sigma \in X(\ell)}[{\norm{f_\sigma}}\norm{f^\sigma}] + \E_{\sigma \in \subheavy_\ell(f)}[{\norm{f_\sigma}}]\!\cdot\!\norm{\heavy_\ell(f)} \\[5pt]&\le
    \frac{\ell+2}{\beta_\ell}\left(\E_{\sigma \in X(\ell-1)}[{\norm{f_\sigma}^2}] + \lambda\norm{f}\right) + \E_{\sigma \in \subheavy_\ell(f)}[{\norm{f_\sigma}}]\!\cdot\!\norm{\heavy_\ell(f)},
    \end{aligned}
    \end{equation}
    where the first inequality holds by the definition of non-heavy faces (for the left summand) and that $\norm{f_\sigma}^2 \le \norm{f_\sigma}$ (for the right summand), the second inequality follows from the law of total expectation, and the last inequality holds by Lemma~\ref{lem-bound-near-times-infront}.

    Applying \eqref{prop-exp-from-non-heavy-faces-eq1} iteratively with $\ell = k-1, k-2, \dots, 0$ yields
    \begin{equation}\label{prop-exp-from-non-heavy-faces-eq2}
    \begin{aligned}
    \E_{\sigma \in X(k-1)}[{\norm{f_\sigma}^2}] \le{}
    &\frac{(k+1)!}{\beta_0\dotsb\beta_{k-1}}\left(\norm{f}^2 + \sum_{\ell=0}^{k-1}\frac{\beta_0\dotsb\beta_{\ell-1}}{(\ell+1)!}\lambda\norm{f}\right) +{} \\&\sum_{\ell=0}^{k-1}\frac{(k+1)!}{\beta_{\ell+1}\dotsb\beta_{k-1}(\ell+2)!}\E_{\sigma \in \subheavy_\ell(f)}[{\norm{f_\sigma}}]\!\cdot\!\norm{\heavy_\ell(f)} \\[5pt]\le{}
    &\frac{(k+1)!}{\beta_0\dotsb\beta_{k-1}}\left(\norm{f} + 2\lambda\right)\norm{f} +{} \\&\sum_{\ell=0}^{k-1}\frac{(k+1)!}{\beta_{\ell+1}\dotsb\beta_{k-1}(\ell+2)!}\E_{\sigma \in \subheavy_\ell(f)}[{\norm{f_\sigma}}]\!\cdot\!\norm{\heavy_\ell(f)} \\[5pt]\le{}
    &\left(\frac{\beta_k}{k+2} - \lambda\right)\norm{f} +{} \\&\sum_{\ell=0}^{k-1}\frac{(k+1)!}{\beta_{\ell+1}\dotsb\beta_{k-1}(\ell+2)!}\E_{\sigma \in \subheavy_\ell(f)}[{\norm{f_\sigma}}]\!\cdot\!\norm{\heavy_\ell(f)},
    \end{aligned}
    \end{equation}
    where the second inequality holds since $\beta_\ell \le 1$ for every $\ell \ge 0$, and the last inequality holds since $\norm{f} \le \big(\frac{\beta_k}{k+2}-\lambda\big)\frac{\beta_0\dotsb\beta_{k-1}}{(k+1)!} - 2\lambda$.
    
    Substituting \eqref{prop-exp-from-non-heavy-faces-eq2} in \eqref{prop-exp-from-non-heavy-faces-eq0} implies
    \begin{equation}\label{prop-exp-from-non-heavy-faces-eq3}
    \E_{\sigma \in X(k)}[\norm{f_\sigma}\norm{f^\sigma}] \le \frac{\beta_k}{k+2}\norm{f} + \sum_{\ell=0}^{k-1}\frac{(k+1)!}{\beta_{\ell+1}\dotsb\beta_{k-1}(\ell+2)!}\E_{\sigma \in \subheavy_\ell(f)}[{\norm{f_\sigma}}]\!\cdot\!\norm{\heavy_\ell(f)}.
    \end{equation}
    
    Finally, substituting~\eqref{prop-exp-from-non-heavy-faces-eq3} in Lemma~\ref{lem-exp-from-non-heavy-faces} completes the proof.    
\end{proof}

\subsubsection*{Proofs of Lemmas~\ref{lem-exp-from-non-heavy-faces} and~\ref{lem-bound-near-times-infront}}

\begin{proof}[Proof of Lemma \ref{lem-exp-from-non-heavy-faces}]
Let $f \in C^k(X,\mathcal{F})$. For ease of notation, we identify $f$ and $\delta f$ with their support, i.e., we write $\sigma \in f$ and $\tau \in \delta f$ when $f(\sigma) \ne 0$ and $\delta f(\tau) \ne 0$, respectively.

Let us denote by $\delta_1(f)$ the set of $(k+1)$-faces in $\delta f$ that contain exactly one $k$-face in $f$. Formally,
$$\delta_1(f) = \big\{\tau \in \delta f \;\big|\; |\{\sigma \subset \tau \mid \sigma \in f\}| = 1\big\}.$$
Note that $\norm{\delta f} \ge \norm{\delta_1(f)}$, since by definition, $\tau \in \delta_1(f)$ implies $\tau \in \delta f$ (i.e., $\delta f(\tau) \ne 0)$). So it is enough to lower bound $\norm{\delta_1(f)}$.

First, for any $\tau \in X(k+1)$ such that $|\{\sigma \subset \tau \mid \sigma \in f\}| = 1$, if $\sigma$ is the only $k$-face in $\tau$ that is in $f$ (i.e., $\sigma$ is the only $k$-face in $\tau$ for which $f(\sigma) \ne 0$), then $\sigma$ is the only $k$-face that contributes to $\delta f(\tau)$. It follows that
\begin{equation}\label{lem-exp-from-non-heavy-faces-eq-1}
    \tau \in \delta f \;\iff\; \tau\setminus\sigma \in \delta(f_\sigma).
\end{equation}

Now, let $\sigma \in f$. Recall the probability distribution over $Y_\sigma^k$ (where $\sigma$ is a $k$-face): 
the probability of a $k$-face $\rho \in Y_\sigma^k$ is the probability that $\rho = (\sigma \cup \{v\}) \setminus \{u\}$, where $v \in X_\sigma(0)$ is chosen according to the probability distribution of $X_\sigma(0)$ and $u \in \sigma$ is chosen uniformly at random. For readability, in what follows we slightly abuse notation and write $\sigma v$ and $\sigma v\hat{u}$ as shorthand for $(\sigma \cup \{v\})$ and $(\sigma \cup \{v\}) \setminus \{u\}$, respectively.
It holds that
\begin{equation}\label{lem-exp-from-non-heavy-faces-eq-2}
\begin{aligned}
    \norm{f^\sigma} &= \Pr_{\substack{v \in X_\sigma(0)\\u \in \sigma}}[\sigma v\hat{u} \in f] \\[5pt]&=
    \frac{1}{k+1}\sum_{u \in \sigma}\Pr_{v \in X_\sigma(0)}[\sigma v\hat{u} \in f] \\[5pt]&\ge
    \frac{1}{k+1}\Pr_{v \in X_\sigma(0)}[\exists u \in \sigma \text{ s.t. } \sigma v \hat{u} \in f].
\end{aligned}
\end{equation}
Hence, by the definition of $\delta_1(f)$,
\begin{equation}\label{lem-exp-from-non-heavy-faces-eq-3}
\begin{aligned}
    \Pr_{v \in X_\sigma(0)}[\sigma v \in \delta_1(f)] &=
    \Pr_{v \in X_\sigma(0)}\left[\sigma v \in \delta f \;\wedge\; |\{\rho \subset \sigma v \mid \rho \in f\}| = 1\right] \\[8pt]&=
    \Pr_{v \in X_\sigma(0)}[\sigma v \in \delta f \;\wedge\; \forall u \in \sigma,\, \sigma v \hat{u} \notin f] \\[8pt]&=
    \Pr_{v \in X_\sigma(0)}[v \in \delta(f_\sigma) \;\wedge\; \forall u \in \sigma,\, \sigma v \hat{u} \notin f] \\[8pt]&\ge
    \Pr_{v \in X_\sigma(0)}[v \in \delta(f_\sigma)] - \Pr_{v \in X_\sigma(0)}[\exists u \in \sigma \text{ s.t. } \sigma v \hat{u} \in f] \\[8pt]&\ge
    \beta_k - (k+1)\norm{f^\sigma},
\end{aligned}
\end{equation}
where the second equality holds since $\sigma \in f$, the third equality follows by~\eqref{lem-exp-from-non-heavy-faces-eq-1}, and the last inequality follows by~\eqref{lem-exp-from-non-heavy-faces-eq-2}, the $\beta_k$-coboundary expansion of $X_\sigma$ in dimension $-1$, and the assumption that $\sigma \in f$, so $\norm{f_\sigma} = 1$.

Now, on one hand,
\begin{equation}\label{lem-exp-from-non-heavy-faces-eq-4}
\begin{aligned}
    \Pr_{\substack{\sigma \in X(k) \\ v \in X_\sigma(0)}}[\sigma \in f \;\wedge\; \sigma v \in \delta_1(f)] &=
    \E_{\sigma \in X(k)}\bigg[\Pr_{v \in X_\sigma(0)}[\sigma \in f \;\wedge\; \sigma v \in \delta_1(f)]\bigg] \\&=
    \E_{\sigma \in X(k)}\bigg[\mathbf{1}_{\{\sigma \in f\}}(\sigma)\cdot\Pr_{v \in X_\sigma(0)}[\sigma v \in \delta_1(f)]\bigg] \\[8pt]&\ge
    \E_{\sigma \in X(k)}\bigg[\mathbf{1}_{\{\sigma \in f\}}(\sigma)\cdot\Big(\beta_k - (k+1)\norm{f^\sigma}\Big)\bigg] \\[8pt]&=
    \E_{\sigma \in X(k)}\bigg[\norm{f_\sigma}\Big(\beta_k - (k+1)\norm{f^\sigma}\Big)\bigg] \\[12pt]&=
    \beta_k\norm{f} - \E_{\sigma \in X(k)}(k+1)\norm{f_\sigma}\norm{f^\sigma},
\end{aligned}
\end{equation}
where $\mathbf{1}_{\{\sigma \in f\}}$ denotes the indicator function for the event $\sigma \in f$; the second equality holds since after fixing $\sigma$, the condition $\sigma \in f$ in the inner probability is always true for $\sigma \in f$ and always false for $\sigma \notin f$, the inequality follows by~\eqref{lem-exp-from-non-heavy-faces-eq-3}, and the third equality holds since $f$ is a $k$-cochain, so $\norm{f_\sigma} = \mathbf{1}_{\{\sigma \in f\}}(\sigma)$.

On the other hand,
\begin{equation}\label{lem-exp-from-non-heavy-faces-eq-5}
\begin{aligned}
    \Pr_{\substack{\sigma \in X(k) \\ v \in X_\sigma(0)}}[\sigma \in f \;\wedge\; \sigma v \in \delta_1(f)] &=
    \Pr_{\substack{\tau \in X(k+1)\\ \tau \supset \sigma \in X(k)}}[\tau \in \delta_1(f) \;\wedge\; \sigma \in f] \\[5pt]&=
    \Pr_{\tau \in X(k+1)}[\tau \in \delta_1(f)]\Pr_{\substack{\tau \in X(k+1)\\ \tau \supset \sigma \in X(k)}}[\sigma \in f \mid \tau \in \delta_1(f)] \\[1pt]&=
    \Pr_{\tau \in X(k+1)}[\tau \in \delta_1(f)]\frac{1}{k+2} \\[10pt]&=
    \frac{1}{k+2}\norm{\delta_1(f)},
\end{aligned}
\end{equation}
where the first equality holds since both probability distributions are equivalent, the second equality is just conditional probability, and the third equality follows from the definition of $\delta_1(f)$.

Finally, combining~\eqref{lem-exp-from-non-heavy-faces-eq-4} and~\eqref{lem-exp-from-non-heavy-faces-eq-5} completes the proof.
\end{proof}

\begin{proof}[Proof of Lemma \ref{lem-bound-near-times-infront}]
Let $f \in C^k(X,\mathcal{F})$. For ease of notation, as in the proof of previous lemma, we identify $f$ with its support, i.e., we write $\sigma \in f$ when $f(\sigma) \ne 0$.

By the definition of $f^\sigma$,
\begin{equation}\label{lem-bound-near-times-infront-eq1}
\E_{\sigma \in X(\ell)}[\norm{f_\sigma}\norm{f^\sigma}] =
\E_{\sigma \in X(\ell)}\E_{u \in \sigma}[\norm{f_\sigma}\|(f_{\sigma\setminus u})^u\|] =
\E_{\sigma \in X(\ell-1)}\E_{u \in X_\sigma(0)}[\norm{f_{\sigma u}}\norm{(f_{\sigma})^u}].
\end{equation}

Let $\sigma \in X(\ell-1)$. Define the following two random walks on the $(k-\ell)$-faces in $X_\sigma$:
\begin{enumerate}
    \item Move from a face $\tau \in X_\sigma(k-\ell)$ to a vertex $u \in \tau$, and then to a face $\tau' \in X_\sigma(k-\ell)$, such that $u \cupdot \tau' \in X_\sigma(k-\ell+1)$.
    \item Move from a face $\tau \in X_\sigma(k-\ell)$ to a vertex $u \in X_\sigma(0)$, such that $u \cupdot \tau \in X_\sigma(k-\ell+1)$, and then to a face $\tau' \in X_\sigma(k-\ell)$, such that $\tau' \ni u$.
\end{enumerate}



Random walk \#1 can be described by the following operators: $A_1: \R^{X_\sigma(k-\ell)} \to \R^{X_\sigma(0)}$ which moves from a $(k-\ell)$-face in $X_\sigma$ to a vertex that is contained in it, and $B_1 : \R^{X_\sigma(0)} \to \R^{X_\sigma(k-\ell)}$ which then moves from a vertex to a $(k-\ell)$-face that completes it to a $(k-\ell+1)$-face.
Similarly, random walk \#2 can be described by the following operators: $A_2: \R^{X_\sigma(k-\ell)} \to \R^{X_\sigma(0)}$ which moves from a $(k-\ell)$-face in $X_\sigma$ to a vertex that completes it to a $(k-\ell+1)$-face, and $B_2 : \R^{X_\sigma(0)} \to \R^{X_\sigma(k-\ell)}$ which then moves from a vertex to a $(k-\ell)$-face that contains it.

Therefore, random walk \#1 is given by applying the operator $B_1A_1 : \R^{X_\sigma(k-\ell)} \to \R^{X_\sigma(k-\ell)}$ and random walk \#2 is given by applying the operator $B_2A_2 : \R^{X_\sigma(k-\ell)} \to \R^{X_\sigma(k-\ell)}$. Note that if we consider the random walks which are defined by the same operators but with opposite order, i.e., $A_1B_1 : \R^{X_\sigma(0)} \to \R^{X_\sigma(0)}$ and $A_2B_2 : \R^{X_\sigma(0)} \to \R^{X_\sigma(0)}$ we get (in both cases) the well-known non-lazy random walk on the underlying graph of $X_\sigma$. For instance, $A_1B_1$ moves from a vertex $u \in X_\sigma(0)$ to a $(k-\ell)$-face $\tau \in X_\sigma(k-\ell)$ such that $u\cupdot\tau \in X_\sigma(k-\ell+1)$ and then to a vertex $v \in \tau$, namely, it moves from $u$ to a random neighbor $v \ne u$.

Since $A_1,B_1,A_2$ and $B_2$ are linear operators, the eigenvalues of $B_1A_1$ and $B_2A_2$ are equal to the eigenvalues of $A_1B_1$ and $A_2B_2$ (up to the multiplicity of the $0$ eigenvalue). Therefore, $\lambda_2(B_1A_1) = \lambda_2(A_1B_1) = \lambda_2(A_2B_2) = \lambda_2(B_2A_2)$, where the equality in the middle is because both $A_1B_1$ and $A_2B_2$ describe the same random walk.

Now, consider the undirected graph $G(X_\sigma) = (V(X_\sigma), E(X_\sigma))$, where $V(X_\sigma)$ corresponds to the $(k-\ell)$-faces of $X_\sigma$ and $E(X_\sigma)$ corresponds to a step in either of the random walks \#1 and \#2, i.e., its random walk operator is given by $\frac{1}{2}(B_1A_1 + B_2A_2)$. Since $X$ is a $\lambda$-one-sided local spectral expander and by the explanation above, it follows that $\lambda_2\big(\frac{1}{2}(B_1A_1 + B_2A_2)\big) = \lambda$.

It follows that
\begin{equation}\label{lem-bound-near-times-infront-eq2}
\begin{aligned}
\E_{u \in X_\sigma(0)}[\norm{f_{\sigma u}}\norm{(f_{\sigma})^u}] &= 
\E_{u \in X_\sigma(0)}\Bigg[\Pr_{\substack{\tau \in X_\sigma(k-\ell) \,:\\ u \in \tau}}[\tau \in f_\sigma] \cdot \Pr_{\substack{\tau' \in X_\sigma(k-\ell) \,: \\u \cupdot \tau' \in X_\sigma(k-\ell+1)}}[\tau' \in f_\sigma]\Bigg] \\[5pt]&=
\Pr_{\{\tau,\tau'\} \in E(X_\sigma)}[\tau \in f_\sigma \;\wedge\; \tau' \in f_\sigma] \\[14pt]&\le
\norm{f_\sigma}^2 + \lambda\norm{f_\sigma},
\end{aligned}
\end{equation}
where the second equality follows by the probability distribution of $E(X_\sigma)$, and the inequality follows by Lemma~\ref{many-outgoing-edges}.

Finally, substituting \eqref{lem-bound-near-times-infront-eq2} in \eqref{lem-bound-near-times-infront-eq1} completes the proof.
\end{proof}

\bibliography{Bibliography}

\end{document}